\documentstyle[12pt,graphicx,aps,epsf]{revtex}

\newcommand {\be} {\begin{equation}}
\newcommand {\bea} {\begin{eqnarray} \nonumber }
\newcommand {\ee} {\end{equation}}
\newcommand {\eea} {\end{eqnarray}}

\begin{document}
\parskip=0cm
\noindent

\title {On the Effects of Changing the Boundary Conditions on the
Ground State of Ising Spin Glasses}
\author{
Enzo Marinari and Giorgio Parisi\\
\small Dipartimento di Fisica, INFM and INFN, Universit\`a di Roma
         {\em La Sapienza},\\
\small P. A. Moro 2, 00185 Roma (Italy)\\
\small e-mail: 
{\tt Enzo.Marinari@roma1.infn.it, 
Giorgio.Parisi@roma1.infn.it}}

\date{April 2000}

\maketitle

\begin{abstract}
We compute and analyze couples of ground states of $3D$ spin glass
systems with the same quenched noise but periodic and anti-periodic
boundary conditions for different lattice sizes.  We discuss the
possible different behaviors of the system, we analyze the average
link overlap, the probability distribution of window overlaps (among
ground states computed with different boundary conditions) and the
spatial overlap and link overlap correlation functions.  We establish
that the picture based on Replica Symmetry Breaking correctly
describes the behavior of $3D$ Spin Glasses.
\end{abstract}


\section{Introduction}

Understanding the structure of the states of three dimensional ($3D$)
Ising spin glasses at finite temperature is a very interesting
problem.  Long time ago a mean field theory for spin glasses has been
constructed and successfully compared with the numerical properties of
a long range model, the Sherrington-Kirkpatrick model, where mean
field theory is exact by definition \cite{RSB}.

In the mean field approach one finds that the free energy landscape is
strongly corrugated, and it is characterized by the presence of many
local minima, which correspond to spin configurations very different
one from the other.  There are many low free energy local minima which
have a total free energy which differs from the ground state free
energy of an amount which is of order $1$; these minima contribute to
the probability distribution of the spins also for arbitrary large
volumes.  In some sense one can say that in the infinite volume there
are many equilibrium states (for a better qualification of the
terminology see \cite{MAPARIRUZU}): these states are locally different
one from the other.  For historical reasons this picture goes under
the name of {\em Replica Symmetry Breaking} (RSB).

Since analytic progress on finite dimensional system is hard, the
study of $3D$ systems has mainly been based on numerical simulations.
Most of the numerical simulations have investigated systems of size up
to $V\equiv L^3=16^{3}$ and temperature values greater or equal than
half of the critical temperature ($T_{c}$): the results are in very
good agreement with the RSB picture.  Using the best techniques that
are available to us today it is impossible to thermalize in a
reasonable amount of time systems of this size at much lower
temperature: because of that at today the very low temperature region
cannot be explored using Monte Carlo simulations.  On the other end a
different class of algorithms exists which allow the determination of
the ground state at zero temperature for systems of comparable size
\cite{RIEGER,MARTIN,YOUNG,MIDDLETON}: it is clear that it would be
very interesting to compare the results of the RSB approach with the
numerical results at zero temperature.

We face however a difficulty: in the RSB approach one computes
quantities at finite temperature in the infinite volume limit, while
if we look first for the ground state in a finite volume and only
later we send the volume to infinity we are exchanging the order of
the two limits.  One can assume in a tentative way that the two limits
can be exchanged, in the sense that in the zero temperature limit the
probability distributions of the free energies of different
equilibrium states becomes the probability distributions of the local
minima of the energy: this has been the point of view of the authors
that have employed this technique in the last months, it is not
incompatible with any of the findings obtained to date, and we will do
the same in the following.

Of course at zero temperature the ground state (i.e. the global
minimum of the energy) of a finite volume system with quenched random
couplings selected under a continuous distribution is unique, so that
interesting information can be extracted for example if we consider
the effects of an external perturbation.

Different choices are possible.  In a recent, very interesting paper,
Palassini and Young \cite{PALYOU} have shown that it is possible to
get useful information about the nature of the low $T$ phase of $3D$
Ising spin glasses with Gaussian distributed couplings by studying the
behavior of the ground state after changing the boundary conditions of
an $L^{3}$ lattice system from periodic ($PBC$) to anti-periodic
($ABC$).  They analyze ground states ($GS$) obtained with the same
realization of the quenched disorder and different boundary conditions
(or with disorder realizations that differ only in some locations) for
$L\le 10 $.  In this paper we perform a more detailed analysis of the
same systems by considering a larger set of observables.  Here we use
a modified genetic algorithm which is more efficient than the original
algorithm and that will be described in details elsewhere \cite{MAPA}.
This modified algorithm has allowed us to study larger systems ($L\le
14$) than in former work.  We arrive at the conclusion that the whole
set of data strongly suggests that the picture based on Replica
Symmetry Breaking is correct in describing the behavior of $3D$ Spin
Glasses.

\section{Some Theoretical Considerations}

\subsection{Four Possibilities\protect\label{FOURP}}

As we have already discussed in the introduction we consider a $3D$
Ising spin glass with quenched couplings assigned under a Gaussian
probability distribution with zero expectation value, and we study the
behavior of the ground state after changing the boundary conditions
($BC$) of a simple cubic lattice system of size $V=L^3$ from periodic
($P$) to anti-periodic ($AP$).  When we do such a change the ground
state is usually modified and in the new ground state some spins are
reversed.  Let us call $\sigma(i)$ and $\tau(i)$ the spins in the
ground states with periodic and anti-periodic boundary conditions
respectively.  The local overlap on site $i$ is defined as 

\begin{equation}
  q(i)\equiv\sigma(i)\ \tau(i)\ .
\end{equation}
The link overlap is defined on the links and it is given
by 

\begin{equation}
q_{l}(i,\mu) \equiv q(i)\ q(i+\mu)\ ,
\end{equation}
where $\mu$ is the direction of the link, that can take $D$ positive
and $D$ negative values.  If two spin configurations differ by a
global reverse of the spins their link overlap is identically equal to
one (while their overlap is equal to $-1$).  We will consider in the
following the case where we flip the boundary conditions in the
direction $x$, while we leave them unchanged in directions $y$ and
$z$.

In a first approximation (neglecting the points at the boundaries, see
next subsection) we can define the interface among the flipped and the
non flipped region as the sets of link where $q_{l}(i,\mu)=-1$.  We
are interested in finding out the geometrical properties of this
interface in the infinite volume limit.  Of course the probability of
finding the interface on a random link is given by 

\begin{equation}
\rho \equiv \frac12 (1-q_{l})\ , 
\end{equation}
where $q_{l}$ is the disorder expectation value of $q_{l}(i,\mu)$,
averaged over sites $i$ and directions $\mu$.

Here we discuss a few possible situations:

\begin{enumerate}
    
\item 
The interface is confined in a region of width $L^{z}$ (with $z<1$):
inside this region the interface may have overhangs.  The interface
density goes to zero as $L^{-\alpha}$ with $\alpha\ge 1-z$.  This is
what happens in ferromagnetic models, both with ordered and disordered
Hamiltonians, where $\alpha=1$, al least in the ordered case. We will
see immediately that this possibility is completely excluded by the
data.
    
\item
The wandering exponent $z$ is equal to one and the interface density
goes to zero as $L^{-\alpha}$. We also assume than in the large volume
limit the interface is a fractal object with fractal dimension
$D_{s}=D-\alpha$ (where the space dimension $D$ is equal to 3).  More
precisely we assume that if we define a continuous coordinate (in the
interval $[0-1]$) as $\frac{i}{L}$, in the infinite volume limit the
interface becomes a fractal object defined on the continuum
characterized by a fractal dimension $D_{s}$.  In other words we are
assuming that the interface is not a multi-fractal (i.e. that it can
be characterized by a single fractal dimensions as usually happens for
fractal objects one defines on a lattice in statistical mechanics),
and so the relation $D_{s}=D-\alpha$ is consistent.
    
\item 
The exponent $\alpha$ is equal to zero and the density $\rho$ goes to
a non zero value (i.e. $q_{l}$ does not go to $1$ in the large volume
limit).  Here the interface is space filling and in the above
described continuum limit, the interface is a dense set of measure
$1$.  We expect that the probability that the interface does not
intersect a region $\cal R$, whose size is proportional to the system
size, goes to zero when the volume goes to infinity.  On the contrary
in the previous case such a probability would be a non trivial
function of $\cal R$. As we shall see later, this possibility is the
one realized in the case of RSB.

\item
There is a last possibility, which is unusual, and we mention for
completeness (although its description will take a disproportionate
amount of space): this is when the wandering exponent $z$ is equal to
one and $\alpha$ is greater than zero, but in continuum limit the
interface does not become a fractal object of dimension
$D_{s}=D-\alpha$, but it becomes a dense set.  This happens for
example if we consider a set of parallel planes at distance
proportional to $L^{\alpha}$ or a set of isolated points at distance
$L^{\frac{\alpha}{D}}$.  In both cases the density goes to zero as
$L^{-\alpha}$, but the resulting object is {\em not} a fractal in the
infinite volume limit.  A similar case happens in three dimensions if
we take a random walk of length ${\cal L}=L^{3-\alpha}$ with $\alpha
<1$.  The system is a fractal of dimension $D_{s}$ up to a distance
$\xi(L)$, while it looks homogeneous at larger distances.

In this case the density-density correlation function will scale as

\begin{equation}
  C(x,L) \equiv \left. \overline{ \rho(x) \rho(0) }\right|_L
  \propto L^{-\alpha} x^{-C_{s}}
  \label{CON}
\end{equation}
in the region $1 \ll x \ll \xi(L)$, where $C_{s}=D-D_{s}$ is the
co-dimension of the fractal (the underlying random walk in our case).
We will denote by the upper bar the average over the quenched
disorder.  On the contrary in the region $\xi(L) \ll x$ the
correlation function will scale as

\begin{equation}
  C(x,L) \propto  L^{-2 \alpha}\ .  
  \label{DIS}
\end{equation}
The condition that at large distances the term in equation (\ref{DIS})
dominates over the one in equation (\ref{CON}) implies that
$C_{s}>\alpha$.  Moreover the two formulas must be consistent in the
crossover region where $x\approx\xi(L)$. Imposing this condition one
finds that $ L^{-\alpha}
\xi(L)^{-C_{s}} \propto L^{-2 \alpha}$ and  therefore
we obtain that

\begin{equation} 
  \xi(L) \propto
  L^{\omega}\ , \ \ \ \omega = {\alpha \over C_{s}} <1 \ .
\end{equation} 
This case is anomalous in the sense that usual scaling is not valid
and there is a crossover length which increases as a fractional power
of the side of the system.  The usual scaling law (valid in the region
$x=O(L)$)

\begin{equation}
C(x,L) =L^{-2 \alpha} f(x L^{-1}) 
\end{equation} 
is not satisfied and it is replaced by the unusual relation

\begin{equation} 
C(x,L) =L^{- \alpha -C_{s}} f(x
L^{-1}) + A L^{-2 \alpha}g(x L^{-1})\ .
\end{equation}
Only in the case $\alpha=0$ we are in a familiar situation: in this
case the crossover length $\xi$ does not diverge for large $L$ and we
recover the third case of our list.

\end{enumerate}

\subsection{The Effect of Changing Boundaries}

In the replica approach one finds that at zero temperature there are
many different local minima of the Hamiltonian such that even in the
limit where $L$ goes to infinity the difference among the energy of
the ground state and the energy of such minima is of order $1$.  It is
crucial that the among these minima there are some whose site overlap
$q$ and link overlap $q_l$ with the ground state remains different
from $1$ in large volume limit.  It is also important to assume that
for large volumes there exists a function $f$ such that
$q_{l}=f(q^{2})$, in other words that for fixed overlap $q$ the link
overlap $q_{l}$ does not fluctuate when the volume goes to infinity.

When in a $3D$ spin glass the boundary conditions in the $x$ direction
are flipped from positive to negative it is known that the total
energy changes by an amount $\Delta E$, which increases as a power of
$L$.  In a ferromagnet, where the ground state is unique, the ground
state obtained under anti-periodic condition would be locally similar
to the ground state obtained under periodic boundary conditions.  In
other words in any region of side $M$, if the region does not
intersect the interface, which is just a flat surface, the ground
state spin configuration obtained under anti-periodic boundary
conditions will be equal (or equal to the reverse) to the ground state
obtained under periodic boundary conditions.

A similar conclusion holds in the case of a spin glass behaving
according to the predictions of the droplet model \cite{DROPLET},
where the boundary may be a corrugated surface: in this case we expect
that the link overlap among the ground states obtained with $PBC$ and
$ABC$ tends to one in the infinite volume limit.  On the contrary in
the case of a RSB like behavior the ground state obtained under $ABC$
can be locally similar to one of the low energy excited states: in
this case the link overlap among the two ground states can tend to a
value {\em different} from one as $L\to\infty$.

That an excited state may be selected as new ground state when
changing the boundary conditions is strongly suggested by the fact
that the energy difference among ground states obtained with periodic
and anti-periodic boundary conditions increases with $L$, while the
difference among the ground state energy and the excited state energy
at fixed boundary conditions remains fixed.

The nature of the excited state that is selected when we change the
boundary conditions cannot easily derived using simple arguments.  It
is reasonable to assume that in three dimensions, where $\Delta E$
which increases as a power of $L$, the new ground state will be as
different as possible from the old ground state: one expects the
overlap $q$ among the two ground states to vanish in this limit.

The interest in changing the boundary conditions is also due to the
fact that this is also a convenient method to to study the properties
of the excited states. Other methods can be used to clarify similar
questions, but we will not use them here.

\subsection{Gauge Invariance}

In order to change the boundary conditions in the $x$ direction we can
change the sign of all the couplings connecting the plane $x=x_{0}$
with the plane $x=x_{0}+1$.  The choice of $x_{0}$ does not matter:
the ground state energy does not depend on $x_{0}$, and the ground
state obtained after a different choice of the reversed plane
(e.g. $x=x_{1}$) can be obtained starting from the ground state where
anti-periodic boundary conditions have been enforced at $x_{0}$, by
flipping all the spins in the interval $ x_{0} < x \le x_{1}$.

In other words going from periodic to anti-periodic boundary
conditions is a global change that locally is not visible.  At this
end it is convenient to define quantities which are gauge invariant,
in the sense that they do not change when the plane where the
anti-periodic boundary conditions are imposed.

Let us consider a very simple example: the correct definition of
$q_l(i,+\hat{x})$ is

\begin{eqnarray} \nonumber
  q_l(i,+\hat{x})&\equiv&q(i)\  q(i+\hat{x}) \ \ \ \ 
  \mbox{for} \ \ \ x\ne x_{0}, 
  \\ q_l(i,+\hat{x})&\equiv&-q(i)\  q(i+\hat{x}) \ \ \ \ \mbox{for} \ 
  \ \ x=x_{0}\ ,
\end{eqnarray}
where $x$ is the first component of the three dimensional vector $i$.

A more complex case is the definition of the window overlap
\cite{WINDOW} in a box of side $M$ which intersects the plane
$x=x_{0}$. Here the overlap $q_{M}$ can be defined as

\begin{equation}
  \sum_{i \in {\cal A}}q(i) - 
  \sum_{i \in {\cal B}}q(i)
\end{equation}
where $\cal A$ and $\cal B$ are the set of points of the box which are
respectively on the right and on the left of the plane $x=x_{0}$.

The advantage of this prescription (which can be easily generalized to
more complex situations) is that in absence of a magnetic field
nothing depends on the actual position of the plane where the boundary
conditions are enforced: the fact that the reversal of the spins is
imposed at a given value of $x$ is immaterial.  The presence of this
translational invariance is very effective from a practical point of
view: the expectation value of $q_{l}$ does not depend on the point,
so that data can be taken on the whole lattice. The possibility of
doing measurements on the whole lattice and not only far from the
point where the boundary conditions are changed reduces the
statistical errors in a substantial way.

\section{A First Look to the Numerical Results}

We have computed $16000$ couples of ground states on three dimensional
lattices of linear size $L=4$, $73740$ for $L=6$, $8532$ for $L=8$,
$2434$ for $L=10$, $910$ for $L=12$ and $670$ for $L=14$: for each
sample of the quenched disorder (Gaussian couplings) we have computed the two ground states
under periodic and anti-periodic boundary conditions, using a modified
genetic algorithm which will be described elsewhere \cite{MAPA}.

\subsection{The Average Link Overlap}

The introduction of anti-periodic boundary conditions breaks the
discrete rotational invariance of the simple cubic lattice, so that
the expectation value of $q_{l}(i,\hat{x})$, i.e. of the link overlap
in the $x$ direction, can be different from that of the link overlap
in the other two directions.  In the ferromagnetic case (without
disorder) the change of the boundary conditions produces an interface
in one of the $y-z$ planes.  We will call {\em perpendicular link
overlap} $q_{P}$ the expectation value of $q_{l}(i,\hat{x})$ averaged
over the lattice sites, since it is perpendicular to the $y-z$
interface, while we will call {\em transverse link overlap} $q_{T}$
the average over sites of the quantity $\frac12 \left(
q_{l}(i,\hat{y}) + q_{l}(i,\hat{z}) \right)$.

\begin{figure}
\centering
\includegraphics[width=0.5\textwidth,angle=270]{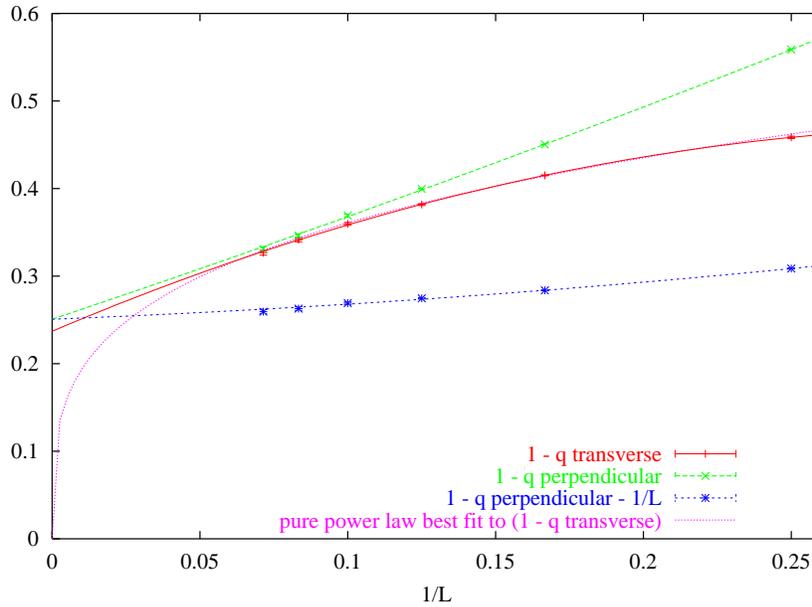}
\caption[a]{$1-q_{T}$, $1-q_{P}$ and $1-q_{P}-\frac{1}{L}$ versus
$\frac{1}{L}$ for $L$ $=$ $4$, $6$, $8$, $10$, $12$, $14$.  We plot
three polynomial best fits (of second degree in $\frac{1}{L}$) to
these three quantities and a simple power best fit to $1-q_{T}$.
\protect\label{qlink} }
\end{figure}

In the ferromagnetic case (without disorder) one finds that

\begin{equation}
 1-q_{P}=\frac{1}{L}, \ \ \ \ 1-q_{T}=0\ .
\end{equation}
We show in figure (\ref{qlink}) the numerical data for $1-q_{T}$ and
$1-q_{P}$ in our $3D$ spin glass.  It is clear that the difference
$q_{T}-q_{P}$ goes to zero with $\frac{1}{L}$: it can be very well
fitted as $a L^{-b}$ with $b\simeq 2.5$ and a normalized $\chi^2$ of
order one.

It is also evident that $1-q_{P}$ is with a very good approximation a
linear function of $L^{-1}$ in the same way it would be in the
ferromagnetic case, with the difference that here the extrapolation in
the $L\to\infty$ limit is clearly different from zero.

A closer look shows that $1-q_{P}$ is not exactly a linear function of
$L^{-1}$, and it can be very well fitted (with a normalized $\chi^{2}$
of order one) as a second order polynomial in $L^{-1}$.  The same
polynomial fit works very well also for $1-q_{T}$.  The two fits
extrapolate to the same value (in the limit given by the statistical
error).  In the same picture we show also the data for
$1-q_{P}-\frac1L$, i.e. the value we measure for $1-q_{P}$ in our spin
glass minus its {\em a priori} lower bound, which coincides with the
value in the ferromagnetic case, and we note it has a very weak
dependence over $L$.

A power law fit to a zero asymptotic value (i.e. to the form $A
L^{-\alpha}$) of $1-q_{P}$ does not give acceptable results, while the
data for $1-q_{T}$ can be fitted by a pure power law, with $\alpha
\simeq 0.27$. A power law fits also works (even if with a higher value
of $\chi^2$) for $1-q_{l}(L)\equiv 1- \frac23 q_{T} -\frac13 q_{P}$
with a similar value of $\alpha$.

It clear that it is not acceptable to fit the values of the two link
overlaps with two different functional forms.  Our conclusion is that
the data do not support the droplet model prediction that $q_{l}$ is
zero in the large $L$ asymptotic limit, and they hint for 

\begin{equation}
  \lim_{L\to\infty}\left(1 - q_{l}\right) 
  = 0.245 \pm 0.015\ .
\label{E-TWOFOURFIVE}
\end{equation}
This estimate of the errors relays on the polynomial fit in powers of
$\frac{1}{L}$. Considering a third order polynomial for $q_{T}$ does
not change the situation. If we fit $1-q_{T}$ with a four parameter
function of the form $A + B L^{-\alpha} + C L ^{-2\alpha}$ we find an
extrapolated value of $A = q_{l} = 0.20 \pm 0.06$, which is consistent
with the previous estimate: it is clear that in providing a complete
estimate of the error over the extrapolated value one has to include
the systematic uncertainty due to the possibility of using different
functional forms for the extrapolating function. On the other end
corrections proportional to a power of $\frac{1}{L}$ arise naturally,
as it will be seen in the section on the correlations functions.

The analysis of $q_{T}$ and $q_{P}$ shows that the RSB picture gives a
consistent and satisfactory explanation of the numerical data.  We
will see in the rest of this note that this conclusion is supported by
the analysis of many other quantities.

\subsection{The Hole Distribution}

It is interesting to consider a region of shape $\cal R$ and to define
the probability $P_1({\cal R},L)$ that a box of such a shape does {\em
not} intersect the interface.  In the case of a box of shape $2\times
1 \times 1$ it is clear that $P_1({\cal R},L)$ coincides with
$\frac{1}{2}(1+q_{l})$.

\begin{figure}
\centering
\includegraphics[width=0.5\textwidth,angle=270]{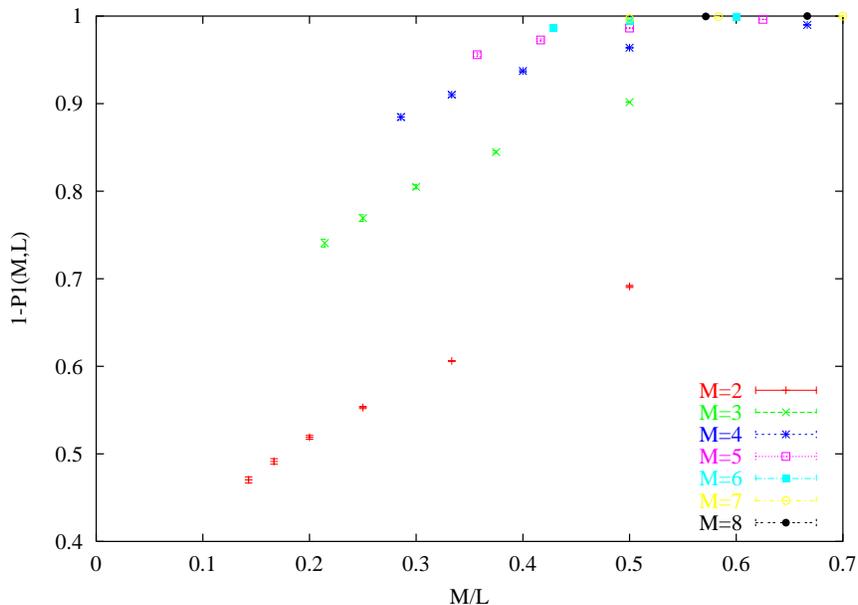}
\caption[a]{ $1-P_1(M,L)$ versus $\frac{M}{L}$ for $M=2, ...,
\min\left(L,8\right)$, and $L$ $=$ $4$, $6$, $8$, $10$, $12$, $14$.
\protect\label{cubi} }
\end{figure}

Following Palassini and Young \cite{PALYOU} we define $1-P_1(M,L)$ as
the probability that the interface does hit a cubic box of side $M$,
i.e. we take ${\cal R} =M\times M \times M$.  In figure (\ref{cubi}) we
show our numerical data for various values of $L$ and $M$. It is
possible to fit these data (if we restrict ourself in the region $M\le
\frac{L}{2}$, in order to have comparable finite volume effects for
the different $L$ values) by a simple power of $L$ (i.e. by the form
$A L^{-\alpha}$): however the power $\alpha$ of the best fits strongly
depends on $M$ (it is given by $0.32$, $0.23$ and $0.14$ for $M$ $=$
$2$, $3$ and $4$ respectively), while in a meaningful fit we would
expect to find  a $M$-independent exponent.

Second order polynomial fits in powers of $\frac{1}{L}$ work well in
the region $M\le \frac{L}{2}$: they give for the extrapolated
$\left(1-P_1\left(M,\infty\right)\right)$ the values $0.35$, $0.56$
and $0.65$ for $M$ $=$ $2$, $3$ and $4$ respectively.  The data are
consistent with the possibility that $1-P_1(M,\infty)$ goes to $1$
when $M\to\infty$.  Indeed the extrapolated data reported before can
be fitted as $1-P_1(M,\infty) = 1 - A M^{-\gamma}$ with $\gamma$ close
to one.

It is clear that also in this case the data do not favor the kind of
behavior predicted by the droplet model, where $1-P_1(M,\infty) =0$: they
are far better consistent with the possibility that $1-P_1(M,\infty)$ is a
non trivial function of $M$, which goes to $1$ (and it is not  zero) when
$M$ goes to infinity.

\begin{figure}
\centering
\includegraphics[width=0.5\textwidth,angle=270]{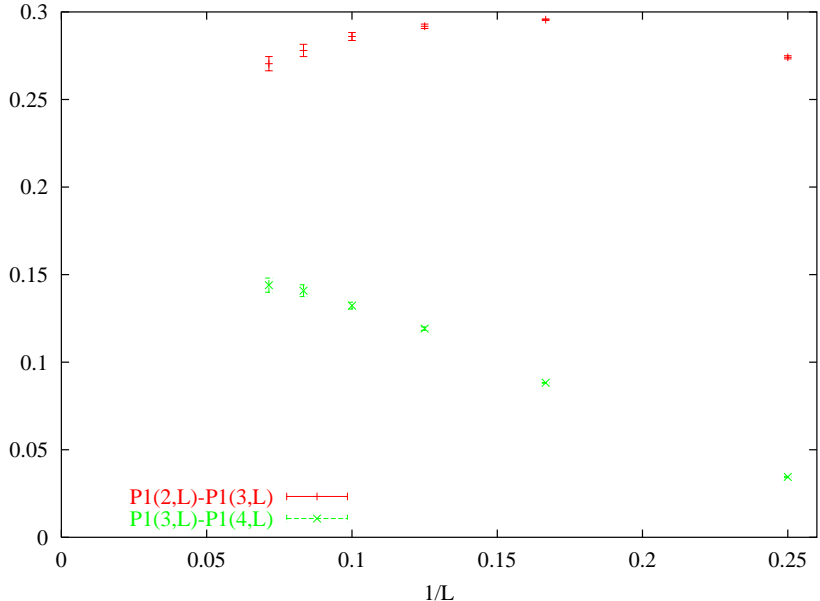}
\caption[a] {The quantities $P_1(2,L)-P_1(3,L)$ and
$P_1(3,L)-P_1(4,L)$ versus $\frac{1}{L}$ for $L$ $=$ $4$, $6$, $8$,
$10$, $12$, $14$.  \protect\label{differenze} }
\end{figure}

This conclusion is strengthened if we look for example to the plot of
$P_1(2,L)-P_1(3,L)$ and $P_1(3,L)-P_1(4,L)$: the data for those two
quantities are shown in figure (\ref{differenze}). The droplet model
suggestion that these quantities go to zero as a power of $L$ when $L$
goes to infinity does not seem very convincing.

\section{Scaling Behavior}

\subsection{The Hole Distribution Again}

\begin{figure}
\centering
\includegraphics[width=0.5\textwidth,angle=270]{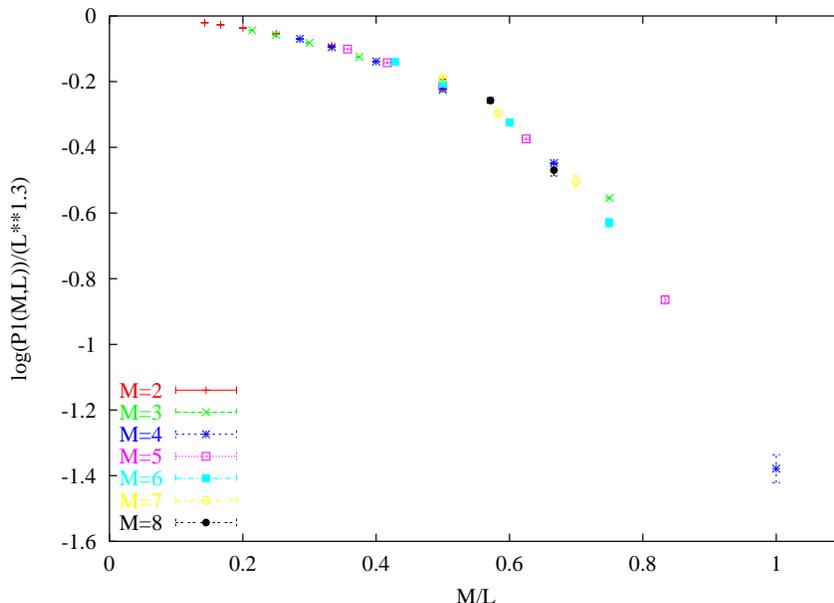}
\caption[a]{$L^{-1.3}\ln(P_1(M,L))$ versus $\frac{M}{L}$ for
$M=2,..., \min\left(L,8\right)$ and $L$ $=$ $4$, $6$, $8$, $10$, $12$, $14$.
\protect\label{cubibis}}
\end{figure}

As we have discussed in the introduction if in the infinite volume
limit the interface becomes a fractal, characterized by an unique
fractal dimension $D_{s}$, and the density of the interface goes to
zero as $L^{-\alpha}$, with $\alpha=D-D_{s}$, the probability of
finding a hole in a box of size $\cal R$ goes to a non trivial
function of $\cal R$ if we keep constant the size of the box in units
of $L$.  In the opposite case the interface would be become a space
filling dense object.  If we consider boxes of side $M$ the previous
argument shows that the probability that the box does intersect the
interface $1-P_1(M,L)$ should be a function of $\frac{M}{L}$.  A
glance to figure (\ref{cubi}) shows that our numerical data do not
exhibit such scaling: the interface is not a fractal characterized by
a simple fractal dimension. On the contrary if we plot
$L^{-1.3}\ln(P_1(M,L))$ versus $\frac{M}{L}$ (see figure
(\ref{cubibis}) we find a reasonable scaling behavior, indicating
again that $1-P_1(M,L)$ goes exponentially to $1$ when $L$ goes to
infinity at fixed $\frac{M}{L}$.

\begin{figure}
\centering
\includegraphics[width=0.5\textwidth,angle=270]{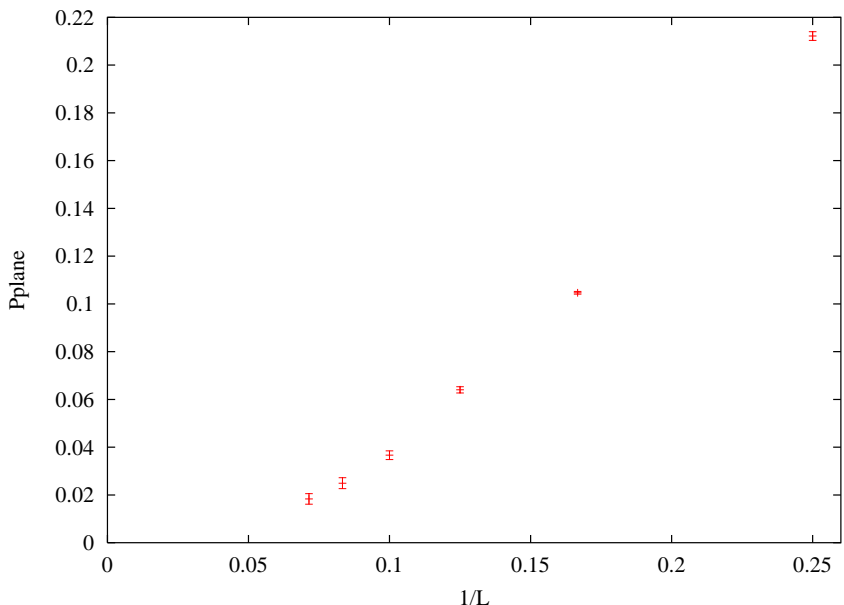}
\caption[a]{$P_{p}(L)$ versus $L^{-1}$ for 
$L$ $=$ $4$, $6$, $8$, $10$, $12$, $14$.  
\protect\label{piani} }
\end{figure}

The same conclusion holds very clearly if we consider the case of
boxes of size $L \times L \times 1$, i.e. planes, oriented in the
$y-z$ directions, i.e. parallel to the interface.  The probability
that such a plane does {\em not} intersect the interface $P_{p}(L)$
would tend to $1$ in the situation (1) described in section
(\ref{FOURP}), while it would tend to a non zero number in situation
(2), The numerical data (in figure (\ref{piani})) decrease very
rapidly as a function of $L$ (they can be reasonably fitted as
$aL^{-\delta}$ with $\delta$ of order two, but a better fit has the
form $a\exp(-bL^{c})$: the asymptotic limit is zero in both cases).
The first scenario we have presented in section (\ref{FOURP}) i.e. the
one of an exponent $z<1$, would imply that this probability goes to
one asymptotically: this is certainly not the case.

This set of data shows that the interface is space filling on a scale
proportional to the side $L$ of the system.  Of course one could
suggest that the numerical data are affected by {\em very} strong
corrections to scaling: however it is not clear why this should happen
(the scaling in figure (\ref{cubibis}) is very good).  In order to see if
we are able to detect any trace of these corrections it is convenient
to consider other quantities.

\subsection{Other Scaling Laws}\label{other}

If we go back to our boxes of side $M$ we can define the window
overlap \cite{WINDOW} as the value of the overlap $q$ restricted to
one of these boxes. We can define $P_{M,L}(|q|)$ as the probability
distribution of the absolute value of such window overlap: the
probability is symmetric, so that we can consider only non negative
values of $q$. The probability that the interface does not intersect
the cube of side $M$ is simply given by $P_{M,L}(1)$.

\begin{figure}
\centering
\includegraphics[width=0.5\textwidth,angle=270]{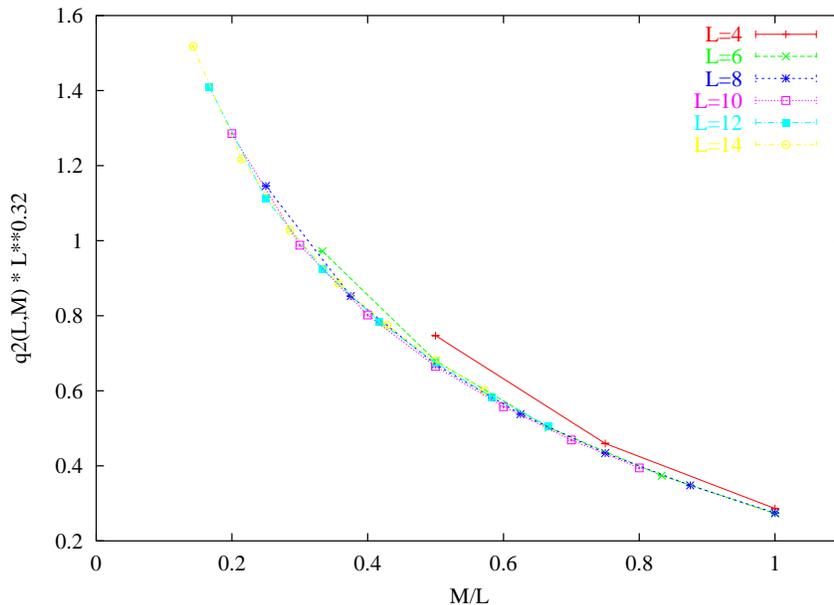}
\caption[a]{ $q^{2}_{M,L}L^{\delta}$ versus
$ML^{-1}$ for $M=2, ..., \min\left(L,8\right)$ and 
$L$ $=$ $4$, $6$, $8$, $10$, $12$, $14$.  
\protect\label{q2cubi} }
\end{figure}

In the case where the two ground states obtained under different
boundary conditions are as different as possible the quantity

\begin{equation}
q^{2}_{M,L} \equiv \int dq\  q^{2}\  P_{M,L}(q)
\end{equation}
should go zero at fixed $\frac{M}{L}$ at large $M$.  A simple possible
scaling behavior is that for large $M$ and $L$

\begin{equation}
  q^{2}_{M,L} \simeq L^{-\delta}f(ML^{-1})\ .
\end{equation}
We show the numerical data for $q^{2}_{M,L}L^{\delta}$ versus
$ML^{-1}$ figure (\ref{q2cubi}) (with $\delta=.32$).  The data for $L>4$
show a beautiful and very accurate scaling behavior.

The data in figures (\ref{cubibis}) and (\ref{q2cubi}) strongly
suggest that asymptotic scaling laws are valid with good accuracy
already for the lattice and block sizes that we have considered:
sub-leading corrections are small, and they do not seem to be of
exceptionally large size.  Moreover we find that for windows that
occupy a finite fraction of the entire system in the infinite volume
limit the window overlap distribution seems to become a delta function
$\delta(q)$.

Up to now all of our numerical evidence seems in perfect agreement
with the RSB predictions, and the first two possibilities we had
discussed in section (\ref{FOURP}) are excluded. In order to exclude the
more exotic last possibility, and to present further evidence for the
correctness of the RSB picture, it is convenient to study the
correlation functions, both of $q_{l}$ and of $q$. This will be done
in the next section.

\section{Correlation Functions}

\subsection {The Link Overlap Correlation Functions}

We will present here a detailed analysis of the correlation functions:
this is important since they carry a large amount of information that
is crucial to distinguish among the different possible behaviors.

As we have already discussed if we consider the case where there is an
interface in the $y-z$ plane we must distinguish among correlations in
the transverse and in the perpendicular directions.  Moreover in the
case of vector like quantities like $q_{l}$ we must distinguish among
correlations in the direction of the link and correlations in the
directions perpendicular to that of the link. For simplicity here we
will consider only one correlation, i.e. the transverse link overlap
correlation in the direction perpendicular to the link.  More
precisely the link overlap correlation function we consider is defined
as

\begin{equation}
  G(d,L) \equiv 
  {1 \over 2 L^{3}}
  \sum_{x,y,z}\overline{q_{l}(x,y,z;\hat{y})\ q_{l}(x,y,z+d;\hat{y})+
  q_{l}(x,y,z;\hat{z})\ q_{l}(x,y+d,z;\hat{z})}\ .
\end{equation}
With this definition the correlation functions are periodic functions
of $d$ with period $L$, and they are symmetric functions of $d$, so
that only the case $0\le d\le \frac{L}{2}$ is interesting.

\begin{figure}
\centering
\includegraphics[width=0.5\textwidth,angle=270]{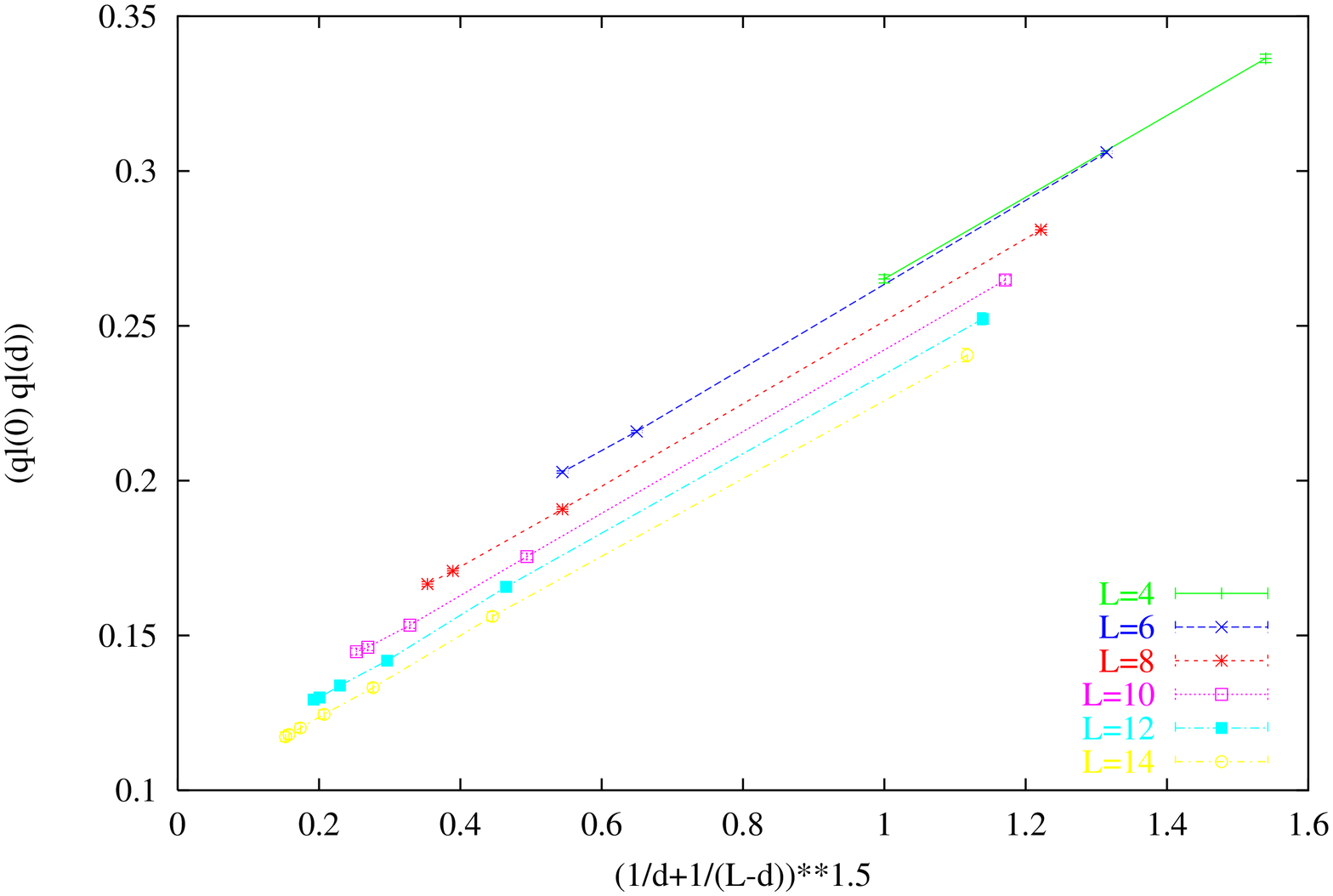}
\caption[a]{The correlation functions $G(d,L)$ versus $w\equiv
\left(\frac{1}{d}+\frac{1}{L-d}\right)^{\lambda}$ with $\lambda=1.5$
for $d> 0$ and $L$ $=$ $4$, $6$, $8$, $10$, $12$, $14$.  The solid
curves are smooth interpolations to the data point, and they turn out
to be remarkably linear and the lines corresponding to different sizes
are nearly parallel.  
\protect\label{G}}
\end{figure}

We show in figure (\ref{G}) the link overlap correlation functions at
distance greater than $0$ for different $L$ values. We plot them as a
function of the variable

\begin{equation}
   w\equiv \left(\frac{1}{d}+\frac{1}{L-d}\right)^{\lambda} \ ,
\end{equation}
with $\lambda=1.5$.  The reasons for this choice of the dependent
variable will be clearer later: right now we just notice that the
points look remarkably linear.

\begin{figure}
\centering
\includegraphics[width=0.5\textwidth,angle=270]{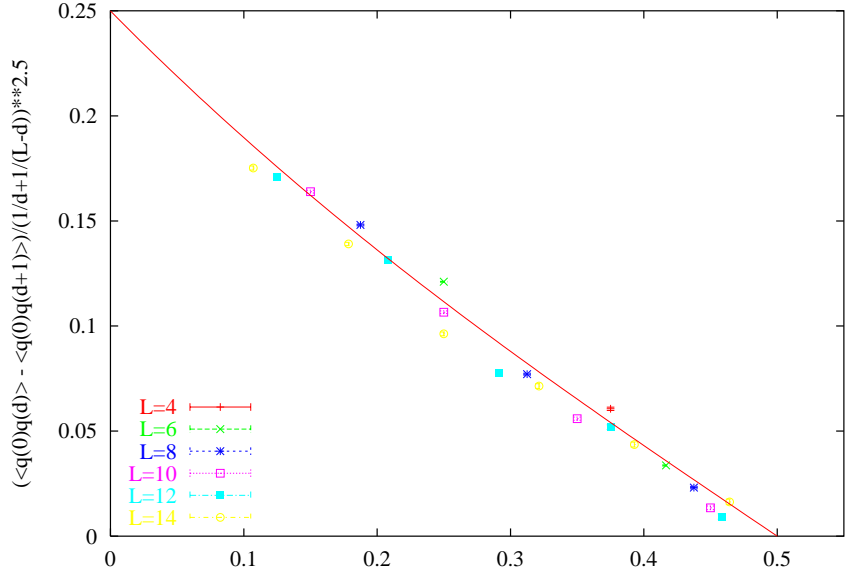}
\caption[a]{The quantity $f(w,L)=\Delta G(d+\frac12,L)
(d^{-1}+(L-d)^{-1})^{1+\lambda}$ with $\lambda=1.5$ as a function of
$d$.  Here $L$ $=$ $4$, $6$, $8$, $10$, $12$, $14$.
\protect\label{DeltaG}}
\end{figure}

If we want to concentrate our attention on the $d$ dependence of the
correlation function (neglecting a possible constant value at large distance) 
we can consider the quantity

\begin{equation}
  \Delta G(d+\frac12,L) \equiv G(d,L)-G(d+1,L)\ .
\end{equation}
Simple scaling implies that the function 

\begin{equation}
f(w,L)\equiv\Delta G(d+\frac12,L) (d^{-1}+(L-d)^{-1})^{1+\lambda}\ ,
\end{equation}
where $w=dL^{-1}$, should be independent from $L$ for large $L$. The
numerical data are shown in figure (\ref{DeltaG}).  A good scaling is
obtained with the choice of the value $\lambda =1.5$.

\begin{figure}
\centering
\includegraphics[width=0.5\textwidth,angle=270]{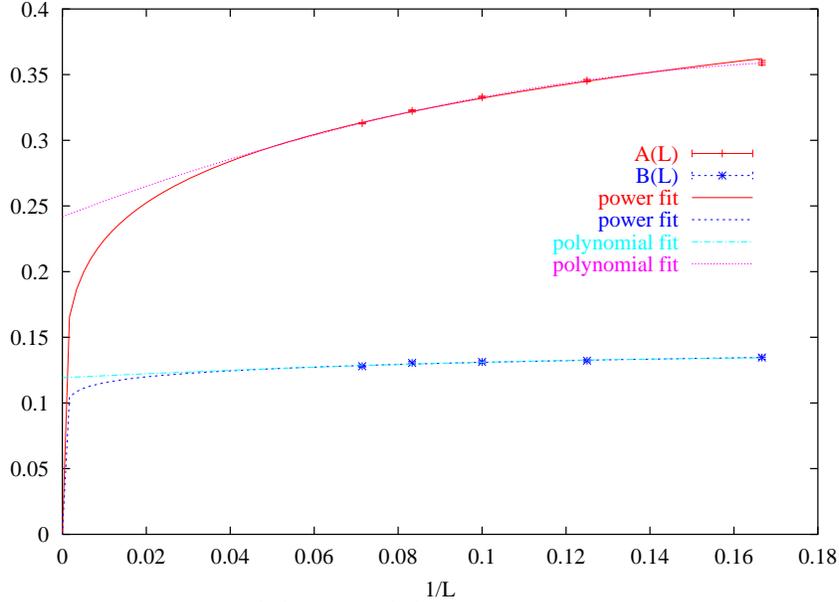}
\caption[a]{The two fit parameters $A(L)$ and $B(L)$ as function of
$L$ for $L$ $=$ $4$, $6$, $8$, $10$, $12$, $14$.  We show both a a
power fit and a polynomial fit in $\frac{1}{L}$.
\protect\label{AB}}
\end{figure}

We expect that the correlation function can be fitted at large $L$ and
$d$ as

\begin{equation}
  G(d,L) = B(L) 
  \left(d^{-1}+(L-d)^{-1}\right)^{-\lambda} 
  + A(L)^{2} \ .
  \label{fitG}
\end{equation}
The form that we have taken is the simplest one which enforces the
periodicity. Of course different forms are possible. 
In order to consider also the $d$-independent part of the correlations
functions it is convenient to fit the whole set of data shown in
figure (\ref{G}) using equation (\ref{fitG}), where for simplicity we
fix $\lambda=1.5$. The fits are good (they would also be good for
similar values of $\lambda$) so that the whole data can be
reconstructed by the knowledge of the parameters $A(L)$ and $B(L)$,
which are shown in figure (\ref{AB}).

While $A(L)$ displays some dependence on $L$, there is no hint that
the quantity $B(L)$ should go to zero when $L\to \infty$.  Certainly
both of them do not go to zero as $L^{-\alpha}$ with $\alpha
\approx.3$ as expected in the case of possibility (4), and they can be
consistently extrapolated to a non-zero value.  If we would insist on
power law fits we would find a value of $\alpha_{A}=0.15$ for $A(L)$
and a ridiculous value of $\alpha_{B}=0.03$ for $B(L)$.

It is remarkable that $A(L)$, that should tend to the link overlap in
the $L\to\infty$ limit, tends to $0.242$, to be compared with the
value $0.245\pm 0.015$ (see equation (\ref{E-TWOFOURFIVE})) that we
have derived from a completely different analysis: the picture that
emerges from these data give consistent support to a RSB physical
scenario.  Also it is worth to comment more on the power fit of figure
(\ref{AB}): even if they are compatible with the data it looks clear
they are only minimally consistent with the physical behavior shown by
the measured points.  Typically these fits stay basically constant for
all measured $L$ values, and than they forecast a sharp descend at
very high $L$ values: the polynomial fits show on the contrary a
consistent behavior in the measured $L$ range and in the $L\to\infty$
limit.

The second scenario we have discussed is in complete variance with
the behavior of the correlation function data.  The exponent $\lambda$
should be $.3$, not $1.5$, as indicated by the data. This conclusion
is in perfect agreement with the results of the hole probability
distribution.

We are now ready to analyze the possible correctness of the fourth
scenario we have discussed (the interface is a dense set in the
continuum limit).  The results of figure (\ref{DeltaG}) show that we
should have $D_{s}=D-\lambda \approx 1.5$.  The fact that the
interface would have a fractal dimension less than 2 is strange, but
not impossible: the interface could look like the surface of
T-lymphocytes, which is full of microvilli, i.e. quasi one dimensional
objects.

However in this case the correlation function (and therefore the
correlation functions at fixed distance) should go to zero as
$L^{-\alpha}$: the correlation functions at distances of order $L$
should go to zero as $L^{-2\alpha}$. The scaling observed in figure
(\ref{DeltaG}) should not be there. The two exponents $\alpha_{A}$ and
$\alpha_{B}$ should be both equal to $0.3$, which we have already seen
it is not true.

It is evident that also the exotic possibility (4) is not compatible
with the observed form of the link overlap correlations (which are
perfectly compatible with simple scaling laws without anomalies), and
the only possibilities is given by the behavior implied by the RSB
scenario.

\subsection {The Overlap Correlation Functions}

Here we consider the overlap correlation function. In this case we can
define two interesting correlation functions: the transverse
correlation function

\begin{equation}
  C_{T}(d,L)={1 \over 2 L^{3}}\sum_{x,y,z}
  \overline{q(x,y,z)\ q(x,y,z+d)+q(x,y,z)\ q(x,y+d,z)}\ ,
\end{equation}
and the perpendicular correlation function

\begin{equation}
  C_{P}(d,L)={1 \over  L^{3}}\sum_{x,y,z}
  \overline{q(x,y,z)\ q(x+d,y,z)\ S(x,x+d)}\ ,
\end{equation}
where $S(x,x+d)=\pm 1 $ is factor which implements gauge invariance:
it is equal to $-1$ if the line which connects the points $x$ and
$x+d$ crosses the plane where boundary conditions are changed.

The two correlations functions are respectively periodic and
anti-periodic functions of $D$ with period $L$ ($C_{P}(L/2,L)=0$).

\begin{figure}
\centering
\includegraphics[width=0.5\textwidth,angle=270]{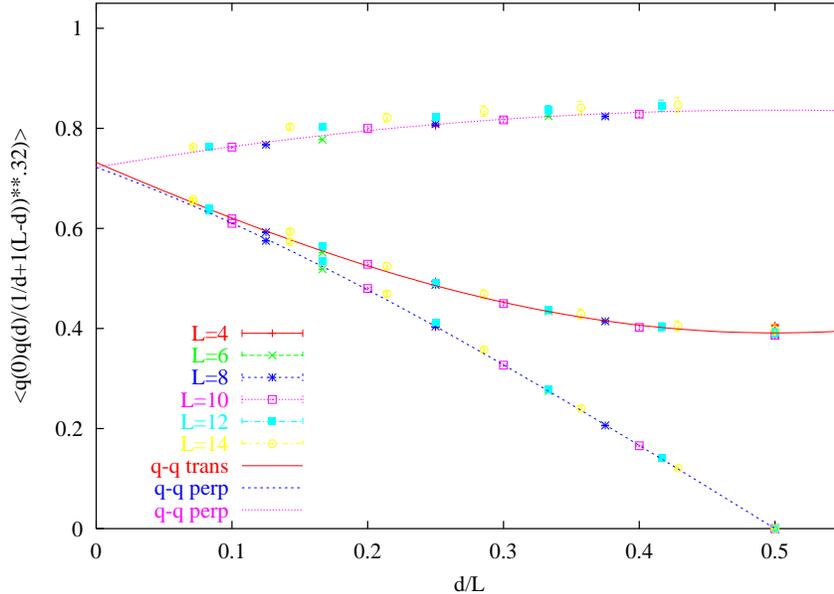}
\caption[a]{The overlap-overlap correlation functions divided times
the factor $\left(\frac{1}{d}+\frac{1}{L-d}\right)^{0.32}$ as a
function of $\frac{d}{L}$. The upper dots are for the perpendicular
correlation function divided an additional factor
$(1-\frac{2d}{L})$. Lines are best fits (see the text).
\protect\label{overcf}}
\end{figure}

We notice that

\begin{equation}
q_{P}(L)=C_{P}(1,L),\ \ \ \ q_{T}(L)=C_{T}(1,L).
\end{equation}
Consequently an analysis of the combined $d$ and $L$ dependence of the
overlap-overlap correlation functions can give information on the
origin of the $L$ dependence of the link overlap, which coincide with
the overlap-overlap correlation function at distance $1$.

According to the previous analysis of the window overlap we expect
that the two correlations functions to scale as

\begin{equation}
  C(d,L)=f(dL^{-1}) (d^{-1}+(L-d)^{-1})^{-\delta}  \ .
\label{fitC}
\end{equation}
The plot of figure (\ref{overcf}) shows a very good scaling, and two
parameter polynomial fits work very well. Transversal and
perpendicular correlation functions go to the same limit when $x\to
0$.  Since the perpendicular correlation function is identically zero
in $d=\frac{L}2$, we also plot the perpendicular $C$ divided times
$(1-\frac{2d}{L})$, removing in this way the main $\frac{d}{L}$
dependence, which comes from a kinematical effect: it goes very
smoothly to the correct limit when $x\to 0$.

The extrapolated values are $0.732 \pm 0.008$ for the transverse
correlation function and $0.722 \pm 0.005$ for the perpendicular one.
This is well consistent with the estimate quoted before (and obtained
by measuring a completely different quantity) for the link overlap,
$q_{l} = 0.755 \pm 0.015$.

\begin{figure}
\centering
\includegraphics[width=0.5\textwidth,angle=270]{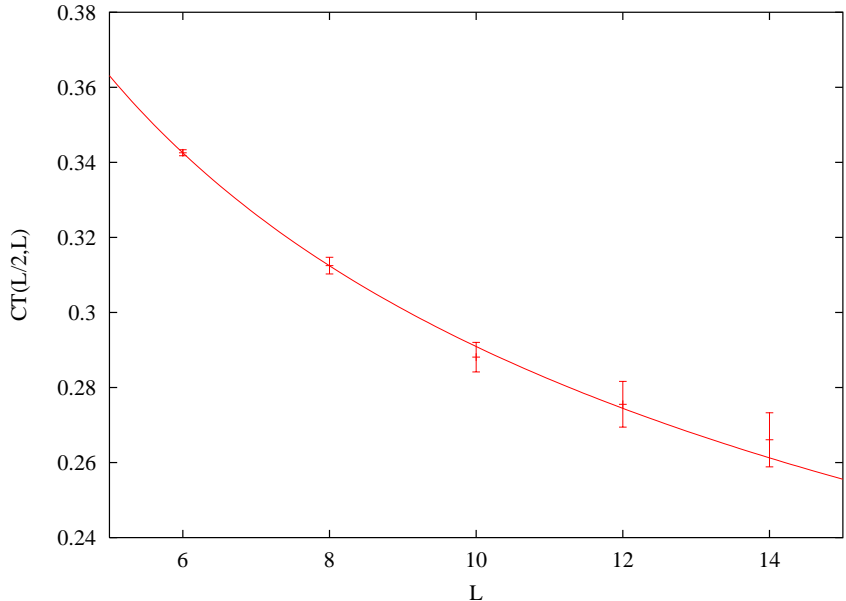}
\caption[a]{The correlation $C_{T}(L/2,L)$ as function  
$L$ $=$ $4$, $6$, $8$, $10$, $12$, $14$.  
\protect\label{meta} }
\end{figure}

As a consistency check we plot in figure (\ref{meta}) the correlation
function $C_{T}(L/2,L)$ as function of $L$.  The data can be fitted
very well by simple power fit $C\ L^{-\delta}$, with $\delta \simeq
0.32$.

It is clear that we have identified a physical mechanism which
naturally generates an $\frac{1}{L}$ dependence in $q_{l}$
(correlation functions do usually feel the size of the system): simple
scaling laws are valid and the ground state structure is non
trivial. On the contrary a trivial structure (in the sense of the
droplet model) would imply that the quantities $f(s)$ extrapolate at
$1$ when $s$ goes to zero (which is evidently not true) or they do not
satisfy a simple scaling and there is a crossover region of length
$\xi(L)$ which goes to infinity with $L$. Both these possibilities are
not compatible with the data which strongly suggested the RSB
scenario.

\section{Conclusions}

We have studied the overlap among the two ground states obtained under
periodic and anti-periodic boundary conditions for $3D$ lattice of
linear sizes $L$ up to $L=14$.  We have analyzed many different
observables: the average of the link overlap, the probability that the
a cube of size $M$ or a plane hits the interface, the window overlap
probability distribution and the correlation functions of the link
overlap and the overlap.

The droplet model in the usual form assumes that the interface density
(which is one minus the link overlap) goes to zero for large $L$ as
$L^{-\alpha}$, and that the interface is a fractal with dimension
$D^{s}=D-\alpha=3-\alpha$.  A small subset of our numerical data is
consistent with the possibility the interface density could go to zero
as $L^{-\alpha}$, but other data cannot be fitted as a simple power
and strongly suggest that the interface density remains finite in the
limit $L \to \infty$.  The possibility that $D_{s}=3-\alpha$ is
completely excluded by the data. Some of the data are compatible with
$\alpha=0.3$ and $D_{s}=1.5$, but we have shown that this exotic
possibility is not compatible with the rest of data.

The RSB scenario is perfectly compatible with the whole set of data.
The appropriate scaling laws for all the variables are satisfied with
amazing accuracy.  The observed $L$ dependence of the interface
density can be easily explained if we consider the scaling law
appropriate for the overlap overlap correlation function.  We also
want to notice the coherence of these results and the recent detection
of large scale excitations in spin glass ground states
\cite{KRZMAR,PALYOU2}.

The behavior we find connects smoothly with the results obtained by
simulations at finite temperature \cite{MAPARIRUZU}.  For example we
find that the overlap-overlap correlation decays as the distance to a
power $\delta=.3$, which is not very far from the value obtained from
simulation at finite temperature.  Other results in this direction can
be found in \cite{MAPARU}.

We think that we have conclusively shown that the ground state
structure of $3D$ Ising spin glasses with Gaussian quenched random
couplings is not trivial.

\section*{Acknowledgments}
We are grateful to M. Palassini and P. Young for a very useful
correspondence.


\begin{references}

\bibitem{RSB} 
  G. Parisi, 
  Phys.  Rev.  Lett.  {\bf 43}, 1754 (1979);
  J. Phys.  A {\bf 13}, 1101, 1887, L115 (1980); 
  Phys.  Rev.  Lett. {\bf 50}, 1946 (1983); 
  M. M\'ezard, G. Parisi and M. A. Virasoro, 
  {\em Spin Glass Theory and Beyond}  
  (World Scientific, Singapore 1987).

\bibitem{MAPARIRUZU} 
  E. Marinari, G. Parisi, F. Ricci Tersenghi, 
  J.J. Ruiz Lorenzo and F. Zuliani,
  J. Stat. Phys. {\bf 98}, 973 (2000),
  cond-mat/9906076.


\bibitem{RIEGER}
  H. Rieger, 
  {\em Frustrated Systems: Ground State Properties via
  Combinatorial Optimization}, in 
  {\em Lecture Notes in Physics} {\bf 501} (Springer-Verlag,
  Heidelberg 1998).

\bibitem{MARTIN}
  J. Houdayer and O. Martin,
  Phys. Rev. Lett. {\bf 83}, 1030 (1999),
  cond-mat/9901276.

\bibitem{YOUNG}
  M. Palassini and A. P. Young,
  Phys. Rev. B {\bf 60}, R9919 (1999),
  cond-mat/9904206.

\bibitem{MIDDLETON}
  A. Middleton, 
  Phys. Rev. Lett. {\bf 83}, 1672 (1999),
  cond-mat/9904285.

\bibitem{PALYOU} 
  M. Palassini and A. P. Young,
  Phys. Rev. Lett. {\bf 83}, 5126 (1999),
  cond-mat/9906323.

\bibitem{MAPA} 
  E. Marinari and G. Parisi,
  to be published.

\bibitem{DROPLET}
  W. L. McMillan,
  J. Phys. C {\bf 17}, 3179 (1984);
  A. J. Bray and M. A. Moore,
  in {\em Heidelberg Colloquium on Glassy Dynamics and Optimization},
  L. Van Hemmen and I. Morgenstern eds. 
  (Springer-Verlag, Heidelberg 1986);
  D. S. Fisher and D. A. Huse,
  Phys. Rev. B {\bf 38}, 386 (1988).

\bibitem{WINDOW}
  E. Marinari, G. Parisi, F. Ricci-Tersenghi and J. J. Ruiz-Lorenzo,
  J. Phys. A {\bf 31}, L481 (1998).

\bibitem{KRZMAR}
  F. Krzakala and O. C. Martin,
  cond-mat/0002055.

\bibitem{PALYOU2} 
  M. Palassini and A. P. Young,
  cond-mat/0002134.

\bibitem{MAPARU}
  E. Marinari, G. Parisi and J. J. Ruiz-Lorenzo,
  to be published.

\end{references}
\end{document}